# Structured Illumination Microscopy using Digital Micro-mirror Device and Coherent Light Source

**Meiqi Li (李美琪)[1] †, Yaning Li (李雅宁)[1] †, Wenhui Liu(刘文辉) [2],**

**Karl Zhanghao(张昊) [1] *, and Peng Xi(席鹏) [1]\*\***

[1] Department of Biomedical Engineering, College of Engineering, Peking University, Beijing 100871, China

[2] Department of Automation, Tsinghua University, Beijing 100084, China

† These authors contributed equally to this work

*Corresponding author: karl.hao.zhang@gmail.com; ** corresponding author: xipeng@pku.edu.cn

Structured illumination microscopy (SIM) achieves doubled spatial resolution by exciting the specimen with a high-contrast, high-frequency sinusoidal pattern. Such an excitation pattern can be generated by interference between multiple laser beams, which are diffracted from a grating. In SIM, 2D imaging requires 9 patterns and 3D imaging requires 15 patterns. Compared to mechanical movement of gratings, opti-electro devices provide rapid switch of the excitation patterns, in which Digital Micro-mirror Device (DMD) is most common in industry. Here we model DMD as the blazed grating and report a fast and cost-efficient SIM. Our home-built DMD-ISIM system reveals the nuclear pore complex and microtubule in mammalian cells with doubled spatial resolution. We further proposed multi-color DMD-ISIM system with simulation, which could potentially exploit the full power of DMD-ISIM.

Keywords: *structured illumination microscopy, super resolution, blazing grating*

OCIS Codes: 180.2520 , 170.0110, 050.1940, 110.3175.

Biologists favor Structured illumination microscopy (SIM) for its high resolution, fast imaging speed, and low photo-toxicity [1-4]. Unlike other super-resolution techniques [5] such as Stimulated Emission Depletion (STED)[6-8] and Single Molecule localization Microscopy (SMLM)[9-11], SIM is compatible with samples prepared for conventional fluorescence microscopy without requiring special sample labeling effort[12, 13]. Recent studies successfully resolve the ultrastructure of cellular organelles and capture their dynamics with SIM[14-17]. Furthermore, a recent work exploits the polarization nature of SIM and achieves super-resolution dipole imaging on existing SIM systems[18]. The technique of polarized SIM (pSIM) was demonstrated on multiple biological systems and revealed "side-by-side" assembly of actin-ring structure in neuronal axons.

The key to obtaining super resolution in SIM is to excite the specimen with high-contrast, fine illumination patterns, which brings high-frequency information into low-frequency region in reciprocal space. To fully cover the doubled region in reciprocal space, 2D-SIM requires 9 illumination patterns (3 directions multiple by 3 phase shifts for each direction); 3D-SIM requires 15 illumination patterns (3 directions multiple by 5 phase shifts for each direction). In early studies, mechanical rotation and translation of grating switches between different patterns, which is quite slow and limits the imaging speed. Afterward, opti-electro device such as liquid crystal spatial light modulator (LC-SLM) [19-22] is adopted into SIM system to achieve fast pattern switching. The imaging speed achieved ~100 reconstructed frames per second and captured the fast dynamics of ER[17], mitochondria[14], and their interactions[15].

Digital Micro-mirror Device (DMD) is one of the most common opti-electro devices in industry. It provides higher pattern refresh frequency and is offered at lower price compared to SLM. Nevertheless, the use of DMD is seldom exploited in SIM. Dan et al. [23-26] reported the first DMD-SIM system which projects the DMD pattern illuminated by incoherent LED light source onto the specimen, which significantly reduces the cost of the SIM system. Nevertheless, the projected incoherent light cannot produce high-contrast illumination patterns [27]. An intuitive realization of DMD-SIM is to simply replace SLM with DMD in the laser-interference-based setup[28]. The obstacle preventing such implementation is the nature of DMD as a blazed grating. The micro-mirror is rotated 12°/-12° in the ON/OFF state so that the surface reflection of the grating is shifted from $0^{th}$ order of the diffraction pattern. The grating is blazed only when the blazing angle, spacing, wavelength, and incident angle cooperates with each other. The blazing angle and spacing are determined by the DMD, so that the DMD-SIM should choose the incident angle and wavelength with special caution.

In this work, we demonstrate that the incident laser must meet the blazing criteria of the DMD to obtain high-contrast illumination pattern. To simplify the problem, we only consider 2D-SIM here while it is also true for 3D-SIM. We start with the simulation of the ±1 order light intensities generated by DMD diffraction, to assist the design of the SIM system with high contrast interference. Furthermore, we built a DMD-based laser interference structured illumination microscopy (DMD-ISIM) system using a single-wavelength laser source (CNI, MGL-FN-561nm-200mW). The measured diffractive pattern from DMD is in accordance with simulation results. In the end, our DMD-ISIM system imaged both standard specimens of fluorescent beads and biological specimens such as nuclear pore complex (NPC) and microtubule with half price and the same frame rate of the SLM-SIM system. The FRC analysis demonstrated doubled spatial resolution in our system.

Multi-color SIM plays a vital role in study intracellular organelle interactions. However, the blazed wavelength is already determined in DMD-ISIM system. Our first effort to achieve multi-color SIM imaging is to find the wavelengths which blaze at the same incident angle. Afterward, we take the ON state and OFF state of DMD as two different blazed gratings, to enable even more imaging colors. Our proposed DMD-ISIM can potentially perform four-color imaging, which matches the performance of existing SIM system, yet at very cost-efficient configuration.

DMD is a kind of spatial light modulator consisting of a micro-mirror array on a CMOS memory unit[29-31]. Each micro-mirror tilted with a given angle $\theta$ along the rotation axis which is at 45 degrees concerning the axis of the square lattice of the mirrors, just as Fig. 1a shows. The deflection of each micro-mirror can be controlled separately by loading different voltage signals to each memory unit. Then the deviation angle of the exit light is changed to realize the modulation of the incident beam. The basic structure of DMD is shown in Fig. 1a. Each micro-mirror is controlled in the binary form corresponding to three states: ON state, OFF state and flat state. When the micro-mirror is in the ON state, the incident light will be deflected into the subsequent light path; when the micro-mirror is on the OFF state, the incident light is deflected away from the subsequent light path; the micro-mirror is flat only when the DMD is powered off.

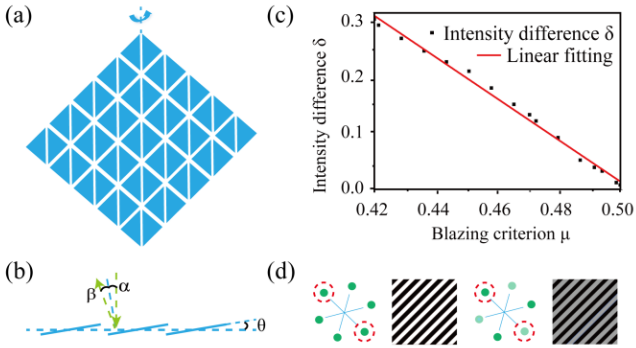

Fig. 1. The simulation results of ±1 order diffractive intensity distribution of DMD. (a) The basic structure of DMD. (b) The illustration of the propagation of light on a single micro-mirror. (c) The relationship between the intensity difference with the blazing criterion $\mu$. (d) The consistent intensity distribution of ±1 order diffractive light produces high contrast interference pattern (left), while differential light intensity produces low contrast streaks (right).

To generate a high-contrast illumination pattern with interference of two laser beams, the two beads should have identical intensity (Figure 1d). The two beams diffracted by the blazed grating of DMD have identical intensity only when the blazing criteria are met. Given that the excitation of DMD is only with a plane wave having a k-vector orthogonal to the rotation axis of the mirrors, the diffraction effect of DMD on light can be reduced to a one-dimensional problem. For a tilt angle $\theta$ in the direction of the diagonal of the mirrors, we can project $\theta$ along the axis of the side length of the square lattice mirrors $\theta' = \arctan(\tan\theta / \sqrt{2})$. The same projection of incident angle $\alpha$ is $\alpha' = \arctan(\tan\alpha / \sqrt{2})$. In order to obtain high-efficiency diffraction and consistent intensity of ±1 order light, the equivalent incident angle $\alpha'$ and reflection angle $\beta'$ have to satisfy the grating equation:

$$\sin(\alpha') + \sin(\beta') = \frac{m\lambda}{d}, \quad (1)$$

with $\beta' = 2\theta' - \alpha'$.

The $\lambda$ denotes the wavelength of the incident light and d indicates the pixel pitch of DMD or spacing of the grating. We can calculate the value m to quantify the blazing condition. The blazing condition is satisfied when m is close to an integer. When m is an integer plus 0.5, it will be the farthest deviation from the blazing condition. To better qualify the blazing condition, we defined the blazing criterion $\mu$ as[32]

$$\mu = |m\backslash1-0.5| = \left|\left[\frac{d}{\lambda}(\sin(\alpha') + \sin(\beta'))\right]\backslash1-0.5\right| \quad (2)$$

If $\mu$ is close to 0.5, it is at the blazing condition. If $\mu$ is close to zero, it's far away from the blazing condition. The difference intensity of ±1 order beams will influence the contrast of the interference fringes. The relationship between the blazing criterion $\mu$ and the normalized intensity difference $\delta$ is shown in Fig. 1c. The contrast decreases when the intensity of the two beams differ from each other. The light intensity difference within 30% which is acceptable for the experiment corresponds to the $\mu$ value surpassing 0.42.

For a given DMD with a fixed pixel pitch $d$, the relationship between wavelength and incident angle is certain under the blazing condition. Therefore, we need to find an appropriate incident angle at a specific wavelength of excitation so that two consistent interference beams can be generated. Based on this, we choose incident wavelength 473 nm and 561 nm which are standard excitation wavelengths of fluorophores. With the simulation results, we figure out the incident angle 51° satisfies the blazing condition for 473 nm excitation and 12° for 561 nm excitation.

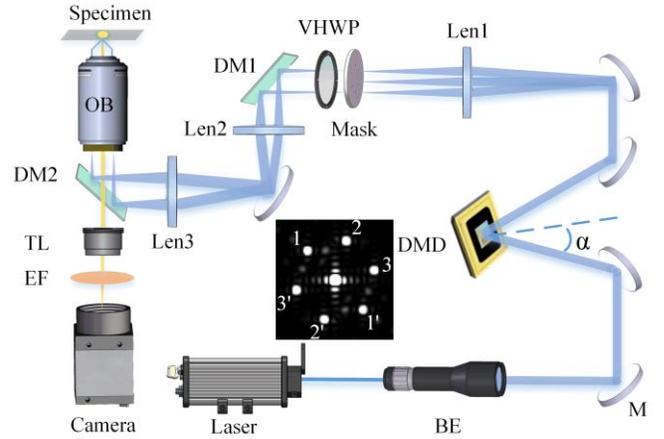

Fig. 2. Schematic of the DMD-ISIM microscopy. BE, beam expander; M, reflecting mirror; DMD, digital micro-mirror device; VHWP, vortex half-wave plate; OB, objective; DM, dichroic mirror; TL, tube lens; EF, emission filter. The inserted image shows the six-beam diffracted light produced by DMD diffraction.

To verify the simulation results, we built up a DMD-ISIM imaging system. The schematic of the system is shown in Fig. 2. The expanded laser beam impinges the DMD at an incident angle $\alpha$ and is diffracted by the DMD (DLP LightCrafter 6500, Texas Instruments). Then, the generated diffraction orders enter lens 1. A mask acts as a spatial filter to block the zero-order beam and preserves the passage of ±1 diffraction orders. The illuminating beam finally comes to the microscope objective through one couple of relay lens (lens2 and lens3) and interference at the objective focal plane. The vortex half-wave plate (VHWP) will change and control the polarization of ±1 order beams to make sure two diffraction beams have the same polarization. Three angles and three phases were realized through the control of each pixel on-off state on DMD. Thus, there are a total of 9 different excitation patterns. The insert image illustrates the ±1 order diffracted light produced by DMD diffraction of three angles. The dichroic mirror DM1 is used to compensate for polarization distortion caused by DM2. Finally, the fluorescent signal is collected by the objective lens and captured by sCMOS (Dhyana 400BSI, TUCSEN) camera through a tube lens.

To ensure the interference fringes formed by two diffraction beams have high contrast, the two diffraction beams should be the same polarization with equal intensity. The polarization direction of two diffraction beams has been controlled by a VHWP to ensure the identical polarization direction. The consistent intensity of two beams is relevant to the blazing condition of DMD. We load diffraction grating patterns with a difference of 120° in three directions into DMD to generate interference fringes. Then we measured the intensity of six diffraction beams after the spatial mask (shown as Fig. 2. marked as 1&1', 2&2' and 3&3'). A comparison of experimental and simulation with normalized diffraction beam intensity in three given conditions are shown in Fig. 3.. When the blazing criterion value is close to 0.5, the two interference beams will have consistent intensity, such as $\lambda=561nm, \theta=12°$ and $\lambda=473nm, \theta=51°$. Otherwise, the intensity of the six beams will fluctuate greatly (Fig. 3c). The experiment results are in good agreement with the simulation calculation.

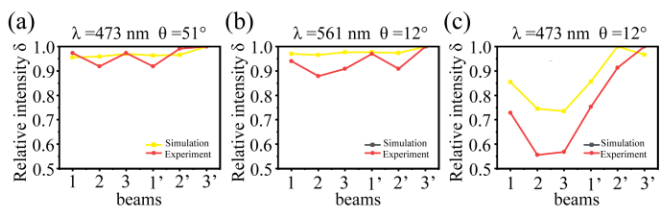

Fig. 3. Experimental verification of DMD diffraction. The intensity distribution of six diffraction beams when the incident angle is 12° with 561nm laser illumination (a) and the incident angle is 51° with 473nm laser illumination (b). (c) The light intensity fluctuates greatly when the blazing condition deviates.

Based on the simulation and experiments analysis, we chose the laser with a central wavelength of 561 nm as the excitation light source in our SIM imaging system. The incident angle of the excitation beam to DMD was set to 12°, which is the condition shown in Fig. 3b. We use 100 nm fluorescent beads to test the resolution of the DMD-ISIM system, as shown in Fig. 4. Since the diameter of the fluorescent beads is smaller than the Abbe diffraction limit, the size of the beads can be neglected during the resolution analysis.

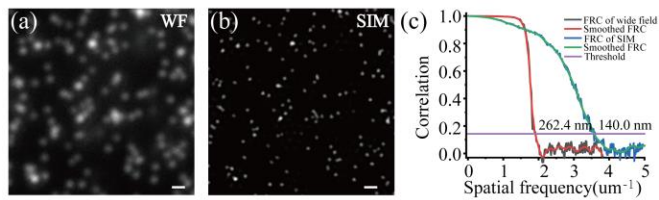

Fig. 4. System resolution test with 100 nm fluorescent beads. (a) WF image of 100 nm fluorescent beads. (b) SIM image of the same region in (a). (c) FRC analysis of images with WF and SIM. Scale bar: 0.5 μm in (a) and (b).

In order to evaluate the resolution better, we use Fourier Ring Correlation (FRC)[33-35] analysis to calculate the resolution of the wide-field (WF) and SIM results, as shown in Fig. 4c. The red line indicates the conventional result with a resolution of 262.4 nm, which is slightly larger than the theory limit 250 nm. The blue line indicates the SIM super-resolution result with a resolution of 140.0 nm, which nearly doubles the resolution of the wide-field result.

To further evaluate the applicability of the DMD-SIM system for biological samples imaging, we image the nuclear pore complex (NPC) and microtubule of COS-7 cells, respectively. The reconstruction is processed with fairSIM [36]. Here, NPC was labeled with Alexa Fluor 555 and tubulin was stained with TMR. As shown in Fig. 5, the resolution of the SIM reconstructed images is significantly higher than that of the wide-field images. Meanwhile, we got normalized intensity profiles (as shown in Fig. 5e) of regions indicated by yellow arrows and red arrows in Fig.5c and 5d. Two microtubule fibers that are blurred by WF image could be separated clearly under SIM.

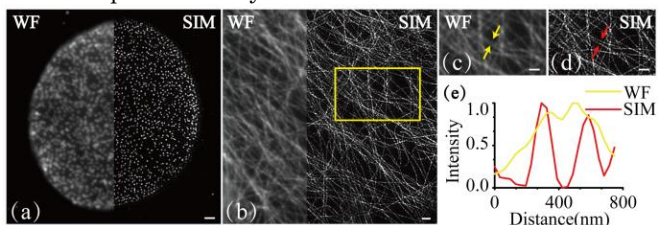

Fig. 5. Super resolution images of biological samples. (a) Image of NPC of COS-7 cell in which the left part is WF result and the right part is SIM result. (b) WF and SIM images of microtubules of COS-7 cell. (c)(d) Magnified view of the yellow area of (b). (e)Normalized intensity profiles along with the marked arrows in (c) and (d). Scale bar, 1 μm in (a), (b) and (c).

The blazed grating is sensitive to wavelength, which limits the realization of multi-color SIM imaging with DMD. To overcome the problem, we simulated the relationship between the blazing criterion with the incident angle and incident wavelength (Fig. 6c). We can see that there are several incident angles satisfying the blazing condition at a particular wavelength. Similarly, there are also multiple incident wavelengths that meet the blazing conditions at a specific incident angle. Fortunately, frequently-used laser lines 375/450/561 nm, 473/640 nm share the same blazing criteria at incident angle 12°. Likewise, laser lines 473/640 also shared the same blazing criteria at the incident angle of 51°. So we can perform two-color and three-color directly under specific choices.

However, if one chooses 561 nm and 640 nm for duel-color imaging, the incident angle has to be different to meet different blazing criteria. In such case, the diffractive beams need to be aligned to enter the subsequent system. Therefore, complicated re-alignment optics are required. Our solution is to take DMD with ON state or OFF state for gratings with different blaze angle. There is 24° rotation (from 12° to -12°) of the micro-mirror between the ON state and OFF state, which allows us to choose an alternate incident angle. For example, we proposed to use the ON state to deflect 375/473/640 nm with incident angle 51° and use the OFF state to deflect 561 nm with incident angle 75°. From Fig. 6c we can see that all the laser wavelength and the incident angle meet the blazing criteria. Although the incident angle of 561 nm laser has a 24° difference compared to 375/473/640 nm laser, the deflected lasers are well aligned because of the same angle difference between the ON/OFF state of DMD (Fig. 6a and 6b). In this way, all the laser lines meet the blazing criteria while keeping the deflected beams aligned. Most importantly, we can achieve four-color imaging with this configuration.

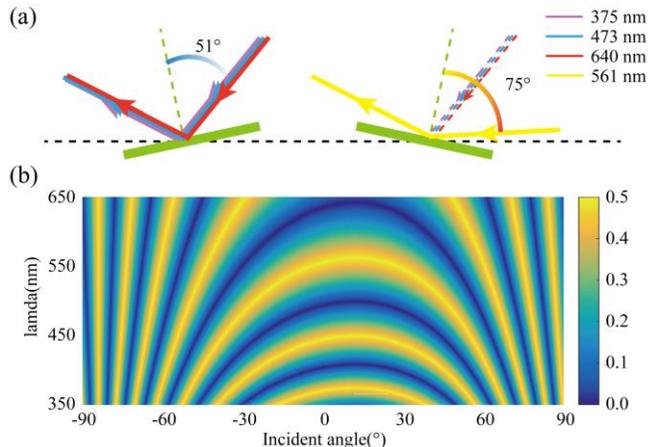

Fig. 6. The simulation and the concept of multi-color imaging. (a) Several wavelengths of specific incident angle 51° satisfied with the blazing condition when micro-mirror is on state. (b) SIM image of the same region in (a). (c) FRC analysis of images with WF and SIM. Scale bar: 0.5 μm in (a) and (b).

In conclusion, we first discuss the effect of intensity uniformity on the DMD-based laser interference SIM system. We further verify the characteristics of DMD blazing diffraction by theory and experiment and give the wavelength and incident angle satisfying the blazed condition. Based on this, we built our homemade DMD-ISIM imaging system and analyzed the resolution of the system. The DMD-SIM system was demonstrated ~2 lateral resolution improvement relative to conventional wide-field microscopy. We further image the subcellular structure with our homebuilt imaging system. The filament structure of microtubules of COS-7 cells can be clearly distinguished under SIM compared with wide-field microscopy. Last but not least, we propose a multi-color SIM imaging system under simulation, which extends the application of DMD-ISIM in research fields such as intracellular organelle interaction.

This work was supported by the National Key Research and Development Program of China (2017YFC0110202), the National Natural Science Foundation of China (61729501, 31971376), the Distinguished Young Scholars of Beijing supported by Beijing Natural Science Foundation (JQ18019), Clinical Medicine Plus X-Young Scholars Project, and Innovative Instrumentation Fund of PKU. KZ acknowledges the support from China Postdoctoral Science Foundation.